\documentclass[amsmath,trackchanges,twocolumn]{aastex702}

\begin{document}


\title{Hard X-ray spectroscopy of a short-period asynchronous polar: IGR J19552+0044}

\author[orcid=0009-0008-1863-8068,sname='Agarwal']{Akash Agarwal}
\affiliation{Astronomy and Astrophysics Group, Raman Research Institute, Bangalore 560 080, India}
\email[show]{akashagarwal@rrimail.rri.res.in}  

\author[orcid=0000-0003-1703-8796,sname='Rana']{Vikram Rana}
\affiliation{Astronomy and Astrophysics Group, Raman Research Institute, Bangalore 560 080, India}
\email[show]{vrana@rri.res.in}

\begin{abstract}

IGR J19552+0044 is an asynchronous polar with a beat modulation of $\approx2$ days due to $\approx3\%$ spin–orbit asynchronism. We present the first hard X-ray properties of this system using a NuSTAR observation. Spectral modeling yields a shock temperature of $kT_s=44.50^{+3.60}_{-3.29}$ keV, implying a white dwarf mass of $M_{\rm WD}\geq 0.85\,M_\odot$. We infer an average accretion rate of $\dot{M} \sim 3.0 \times 10^{-11} \, M_{\odot} \, \mathrm{yr}^{-1}$, consistent with expectations for short-period systems (orbital period $<$ 2 h). Even at this low accretion rate, we detect a significant Compton reflection component that gives the shock height $H\sim 0.3 \, R_{\rm WD}$. Time-resolved spectroscopy of the observation, covering $\approx 61\%$ of a beat cycle, reveals a variation in $\dot{M}$ by a factor of $\sim 4.5$, providing X-ray evidence for the magnetic valve mechanism. Despite the large variations in $\dot{M}$, no significant change in shock temperature is observed. This weak thermal response constrains the shock structure on the white dwarf surface, favoring a shock height that remains approximately constant as $\dot{M}$ changes, potentially due to variations in the fractional accretion area $f$.

\end{abstract}



\section{Introduction} \label{intro}

Cataclysmic Variable stars (CVs) are close binary systems in which a white dwarf (WD) accretes matter from a Roche-lobe-filling late-type companion star \citep{warner1995cataclysmic,hellier2001cataclysmic}. In magnetic CVs (mCVs), the WD magnetic field ($B_{\rm WD}\gtrsim 10^5$ G) disrupts the accretion flow and channels material onto its magnetic poles. Based on field strength, mCVs are broadly classified into two subclasses: Polars and Intermediate Polars (IPs). 
Polars ($B_{\rm WD}\gtrsim 10^7$ G) lack accretion disks and are synchronized systems, with the WD spin period equal to the orbital period \citep{cropper1990polars}. In contrast, IPs ($\sim 10^5$–$10^7$ G) possess truncated disks and usually have a spin period much shorter than the orbital period \citep{patterson1994dq}. A small subset, known as asynchronous polars (APs), exhibits $\sim$1–3\% spin–orbit asynchronism \citep{campbell1999asynchronous}.
This slight asynchronism causes the emission to vary over the beat period, defined as ($(1/P_\omega - 1/P_\Omega)^{-1}$), where $P_\omega$ and $P_\Omega$ are the spin and orbital periods, respectively \citep{hakala2019tess,mason2022magnetic}. 

In mCVs, including APs, the accretion flow from the secondary is funneled along magnetic field lines and forms a strong shock near the magnetic poles on the WD surface \citep{mukai2017x}. In the post-shock region (PSR), the infalling material is cooled primarily through bremsstrahlung emission in X-rays and cyclotron emission at optical wavelengths \citep{cropper1990polars}. The PSR consists of multi-temperature plasma zones as the material slowly cools while settling onto the WD surface \citep{aizu1973x}.  A fraction of the X-rays emitted downward from the PSR is reflected and reprocessed by the WD surface \citep{rothschild1981x}. This reprocessing produces fluorescent emission, most notably the neutral Fe K$\mathrm{\alpha}$ line at 6.4 keV, along with a hump-like excess in the hard X-ray spectrum above 10 keV \citep{ezuka1999iron,mukai2015unambiguous}. The intensity of the reflected emission depends on geometric factors, such as the shock height and the observer’s viewing angle relative to the reflecting region \citep{hayashi2018x}.

CVs evolve from longer ($P_{\Omega} > 3\,\mathrm{h}$) to shorter ($P_{\Omega} < 2\,\mathrm{h}$) orbital periods as they lose angular momentum \citep{paczynski1981gravitational,rappaport1983new}. In short period systems, the mass accretion rate ($\dot{M}$) is typically low ($\dot{M}\sim 10^{-11} M_\odot \text{ yr}^{-1}$)
compared to long orbital period systems \citep{howell2001exploration,knigge2011evolution}. 
At such low accretion rates, cooling in the PSR is typically less efficient (depending on the specific accretion rate), leading to the formation of a shock at a height that is not negligible compared to the white dwarf radius \citep{cropper1999effects,hayashi2014new}. A taller shock height can be inferred observationally when the hard X-ray spectrum does not statistically require a reflection component and the Fe K$\alpha$ line shows a low equivalent width, as seen in EX Hydrae \citep{luna2018constraining}.

IGR J19552+0044 was first identified as an mCV by \citet{masetti2010unveiling} based on optical spectroscopy of hard X-ray sources. Using XMM‑Newton, \citet{bernardini2013nature} studied its X-ray emission, measuring a multitemperature plasma with a maximum temperature of $\sim36$ keV and a low accretion rate. The AP classification was confirmed by \citet{tovmassian2017igr}, who measured photometric and spectroscopic periods of $81.29\pm0.01$ minutes and $83.599\pm0.002$ minutes, corresponding to the WD spin and the binary orbital period, respectively, which imply an asynchronism of 2.8\%. Recent long-term monitoring by \citet{babina2024photometric} clearly revealed a stable, two-peaked brightness modulation with a beat period of 2.0425 days. \citet{tovmassian2017igr} also reported wavelength-dependent variability and cyclotron emission from a magnetic WD with $B<20$ MG. Additionally, \citet{falkenberg2024polarimetry} detected periodically modulated polarized emission, providing further evidence for cyclotron emission from the accretion column.

In this study, we present the first detailed hard X-ray study of IGR J19552+0044 using a NuSTAR observation, with the aim of investigating the Compton reflection and placing constraints on the shock structure above the white dwarf. Section~\ref{secobs} outlines the observations and data reduction procedures. In Section~\ref{secresults}, we present the results of timing, spectral, and time resolved spectral analyses. Section~\ref{secdiscuss} provides a discussion of these findings, and our conclusions are summarized in Section~\ref{secconclu}.

\section{Observations and Data Reduction} \label{secobs}

IGR J19552+0044 was observed by the \textit{NuSTAR} observatory \citep{harrison2013} on 2024 March 24 (ObsID: 30901034002). NuSTAR comprises two co-aligned grazing-incidence telescopes that focus onto focal plane modules FPMA and FPMB, providing highly sensitive observations in the 3–79 keV energy range. Data reduction is done following the standard procedure using HEASOFT-v6.33.2. This involves the use of \texttt{NUSTARDAS} (v2.1.2) along with the calibration database (CALDB) version 20240618. The total duration of the observation is 107 ks. Due to the satellite’s low-Earth orbit, the observation was periodically interrupted by Earth occultations, passages through the South Atlantic Anomaly (SAA), and intervals of high background. The cleaned and calibrated event files were generated using the \texttt{nupipeline} task with the default screening criteria. After filtering out these interruptions, the effective on-source exposure time is 53.4 ks. To maximize the total number of source counts, source spectra were extracted from circular regions with a radius of 70$''$ centered on the source coordinates, while background spectra were taken from nearby source-free regions on the same detector using a radius of 100$''$. Science products, including spectra and light curves, were generated using the \texttt{nuproducts} task, with barycentric correction applied. The resulting spectra were grouped to a minimum of 25 counts per bin using \texttt{grppha} tool to ensure the applicability of the $\chi^2$ statistics for spectral fitting.

\section{Data Analysis and Results} \label{secresults}

\subsection{Timing Analysis} \label{sectiming}

\begin{figure}
	\includegraphics[width=\columnwidth]{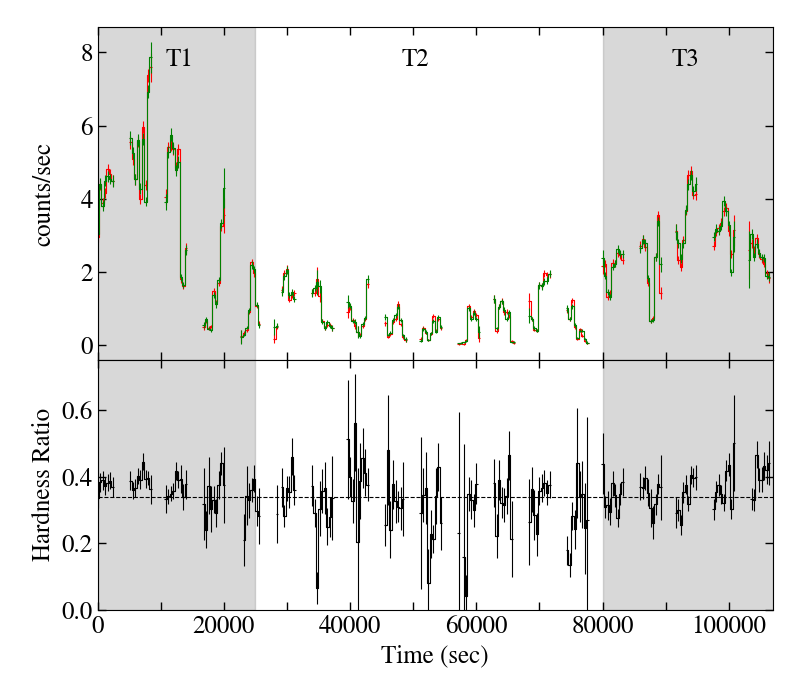}
    \caption{Background-subtracted X-ray light curves and hardness ratio of IGR J19552+0044 observed with NuSTAR. Top: Light curves in the 3–40 keV range from FPMA (red) and FPMB (green), binned at 400 s. Bottom: Hardness ratio, defined as (10–40 keV)/(3–10 keV), for FPMA. The dashed line indicates the mean hardness ratio. Shaded regions mark three intervals: T1, T2, and T3, as referenced in Section \ref{sec:spin_phase_resolved}.}
    \label{fig_lightcurve}
\end{figure}

Background-subtracted light curves from FPMA and FPMB are shown in the top panel of Fig.~\ref{fig_lightcurve}. The analysis was restricted to the 3–40~keV energy range, since the source is dominated by the background at higher energies. The light curve exhibits two broad peaks around 10~ks and 95~ks, indicating a long-term variation with a timescale of $\sim 85$~ks. This timescale is consistent with the clear periodicity detected at optical wavelengths at half of the system’s beat period \citep{babina2024photometric}. The total observation duration of 107 ks therefore corresponds to $\sim 0.61$ of the beat cycle.

To investigate the energy dependence of the variability, we extracted light curves in the 3–10~keV and 10–40~keV bands and defined the hardness ratio as the ratio of their count rates (10–40~keV)/(3–10~keV). As shown in Fig.~\ref{fig_lightcurve} bottom panel, the hardness ratio remains nearly constant, showing no significant spectral change over the observed variability.

We also performed a power spectrum analysis of the background-subtracted FPMA light curve using the CLEAN algorithm \citep{roberts1987time} to identify key periodicities in the system. The light curve was binned at 10~s, and the CLEAN parameters were set to a gain of 0.1 and 1000 iterations. The resulting power spectrum exhibits a broad peak at a period of $\sim 1$ day, consistent with half of the beat period, while no significant power is detected at the expected spin or orbital periods.

\subsection{Spectral analysis} \label{sec:spectral_analysis}

We analyzed the FPMA and FPMB spectra of IGR J19552+0044 using the \textsc{xspec} v12.14.0 software package \citep{arnaud1996xspec}.
 Both spectra were modeled simultaneously to maximize the signal-to-noise ratio, allowing for cross-calibration differences by including a constant normalization factor between them. All uncertainties in the parameters correspond to the 90\% confidence level. 

To account for absorption, we used the \texttt{TBabs} model with the abundance table of \citet{wilms2000absorption} and the photoelectric cross-sections of \citet{verner1996atomic}. 
The X-ray emission from the PSR was initially modeled using a single-temperature plasma model, \texttt{apec}, as this model has been used in studies of many mCVs \citep{mukai2015unambiguous}.
We began the spectral analysis with the model, \texttt{constant} $*$ \texttt{TBabs} $*$ \texttt{apec}, which provided a poor fit to the data, with total $\chi^2/\mathrm{d.o.f.} = 1445/1006$, and showed substantial residuals in the 6–7 keV energy range. To improve the fit, we added a Gaussian component, this provided $\chi^2/\mathrm{d.o.f.} = 1032/1004$. The best-fit plasma temperature was $24.08^{+0.52}_{-0.50}$ keV, representing the average temperature of the X-ray-emitting plasma. The absorption column density was not constrained, and we obtained an upper limit of $< 1.4 \times 10^{21}\,\mathrm{cm}^{-2}$. The best-fit abundance was $Z = 0.17_{-0.17}^{+0.24}$. The line center and width of the \texttt{gaussian} component were \(6.58_{-0.02}^{+0.04}\) keV and \(0.25_{-0.02}^{+0.04}\) keV, respectively.

  \begin{figure}
	\includegraphics[width=\columnwidth]{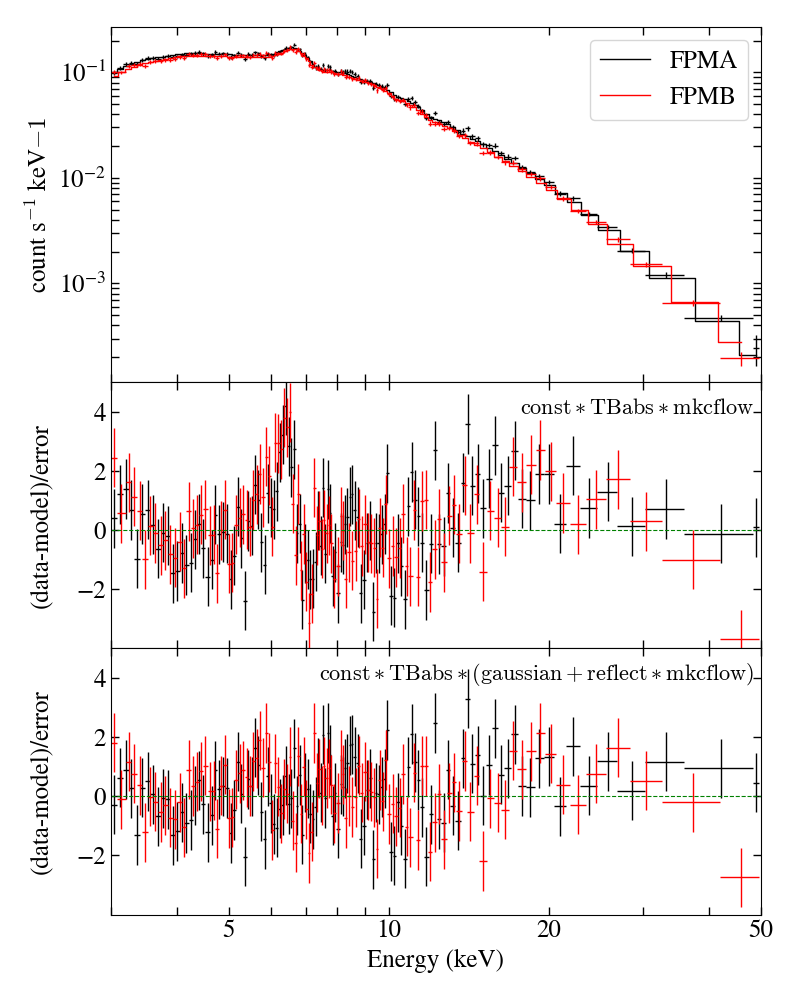}
    \caption{NuSTAR X-ray spectra of IGR J19552+0044 obtained with FPMA (black) and FPMB (red). The top panel shows the observed spectra along with the best-fit model. The middle panel displays the residuals, expressed as (data-model)/error, for an absorbed multi-temperature plasma model. The bottom panel shows the residuals after including gaussian and reflection components, which provide the best fit to the data. For visual clarity, the spectra were rebinned for plotting purposes only.}
    \label{fig:spectrum}
\end{figure}

However, the PSR in mCVs contains a distribution of plasma temperatures \citep{mukai2017x}; therefore, we replaced single temperature \texttt{apec} with \texttt{mkcflow} and adopted the model: \texttt{constant} $*$ \texttt{TBabs} $*$ \texttt{mkcflow}.
The redshift parameter was fixed at $z = 3.885 \times 10^{-8}$, corresponding to a Gaia DR3 distance of $166.5 \pm 1.7$\,pc \citep{2023gaia}.  Due to the lack of data below 3\,keV, we fixed the lowest temperature of the \texttt{mkcflow} component to the minimum allowed value of 0.0808\,keV. This model provided a reasonable fit, with total $\chi^2/\mathrm{d.o.f.} = 1174/1006$ with a maximum plasma temperature of $kT_{\rm max} = 51.07^{+1.83}_{-1.84}$\,keV in the PSR and an absorption column density of $2.46^{+0.39}_{-0.38} \times 10^{22}~\mathrm{cm^{-2}}$. This column density is significantly larger than the galactic column density along the direction of the source $1.4\times 10^{21}~\mathrm{cm^{-2}}$ \citep{bekhti2016hi4pi}. 

The corresponding residuals, shown in the second panel of Fig.~\ref{fig:spectrum}, clearly reveal the reflection characteristics: an iron fluorescence line near 6.4\,keV and a broad reflection hump in the 10–50\,keV range.

\begin{table}
	\centering
	\caption{Best fit model describing the spectra in 3-50 keV. The spectra was modeled using an absorbed, multi-temperature plasma emission model with additional components for iron line emission and reflection \texttt{(const $*$ TBabs(gaussian + reflect(mkcflow)))}. }
    \small
	\label{tab:bestfit}
    \setlength{\tabcolsep}{6pt}
	\begin{tabular}{lccc}
      \hline
    Component & Parameter &   Value \\
      \hline
      Constant          & FPMB              & $1.01\pm0.01$ \\ 
\texttt{TBabs}                   & $N_{\rm H}$ ($10^{22}\,\rm cm^{-2}$) & $1.45^{+0.42}_{-0.41}$ \\ 
\texttt{gaussian}   & LineE (keV)     & $6.43^{+0.07}_{-0.06}$ \\ 
                        & $\sigma$ (keV)     & $0.19\pm0.08$ \\ 
                        & Norm ($10^{-5}$) $^{a}$ & $7.67^{+2.22}_{-1.71}$ \\ 
\texttt{mkcflow}   & $kT_{\mathrm{max}}$ (keV)     & $44.50^{+3.60}_{-3.29}$ \\ 
                        & $Z \, (Z_{\sun})$       & $0.56^{+0.12}_{-0.15}$ \\ 
                        & Norm ($10^{-11}$) $^{b}$ & $4.19\pm0.22$ \\ 
\texttt{reflect}                 & $R_{\rm refl}$  & $0.37^{+0.15}_{-0.13}$ \\ 
\texttt{Flux} & 3-50 keV ($10^{-10}$) $^c$ & $1.03\pm0.01$  \\ 
\hline
\multicolumn{2}{l}{$\chi^2$ (dof)} & $990 (1002)$ \\
\hline
   
\end{tabular}
\begin{flushleft}
\noindent
$^{a}$ \texttt{gaussian} normalization (photons $\mathrm{cm}^{-2}\,\mathrm{s}^{-1}$). \\
$^{b}$ \texttt{mkcflow} normalization ($M_{\odot}\,\mathrm{yr}^{-1}$). \\
$^{c}$ \texttt{Flux} (erg cm$^{-2}$ s$^{-1}$)
\end{flushleft}
\end{table}

To account for the residual excess near 6.4\,keV, we added a \texttt{Gaussian} component to represent the emission due to the neutral Fe K$\alpha$ fluorescence line. This addition significantly improved the spectral fit, giving total $\chi^2/\mathrm{d.o.f.} = 1020/1003$. The center of the Gaussian line was at $E = 6.47^{+0.10}_{-0.07}$\,keV, with an equivalent width $161^{+68}_{-65}$\,eV. With this updated model, the maximum plasma temperature was refined to 
$kT_{\rm max} = 54.78^{+2.53}_{-2.15}$\,keV.

To model 10-50 keV hump feature, we used a Compton reflection component by convolving the \texttt{mkcflow} with the \texttt{reflect} \citep{magdziarz1995angle}. The final model adopted was $\texttt{const}\,*\,\texttt{TBabs}\,*\,(\texttt{gaussian} +\texttt{reflect}\,*\,\texttt{mkcflow} )$. The two abundance parameters of the \texttt{reflect} component (overall abundance and iron abundance) were tied to the abundance parameter of the \texttt{mkcflow} component. and the inclination angle of the reflecting surface was initially fixed at the default value of $\cos\mu = 0.45$ (i.e., $\mu \sim 63^\circ$). As shown in the bottom panel of Fig.~\ref{fig:spectrum}, including the \texttt{gaussian} and \texttt{reflect} components minimizes the residuals around $\sim6.4$ keV and above 10 keV, resulting in a significantly improved fit with \(\chi^2/\mathrm{d.o.f.} = 990/1002\). The reflection amplitude was found to be \(R = 0.37^{+0.15}_{-0.13}\), indicating a moderate level of Compton reflection. Consequently, the maximum plasma temperature decreased to \(kT_{\mathrm{max}} = 44.50^{+3.60}_{-3.29}~\mathrm{keV}\). Other spectral parameters, as summarized in Table~\ref{tab:bestfit}  remained consistent with the parameters fitted without the reflection component, within statistical uncertainties. The absorption column density was found to be \(N_{\rm H}= 1.45^{+0.42}_{-0.41} \times 10^{22}~\mathrm{cm}^{-2}\). This moderate level of absorption is not sufficient to significantly affect the hard X-ray portion of the spectrum and thus does not interfere with or introduce degeneracy in the detection of the reflection component. 

 As discussed by \citet{mukai2015unambiguous}, the reflection amplitude $R$ and the angle $\mu$ are degenerate in reflection modeling. 
 To investigate this degeneracy, we followed the method described by \citet{freitas2026extreme} and performed spectral fits by varying $\cos\mu$ from 0.1 to 0.9 in steps of 0.1. We obtained comparable fit statistics ($\chi^2$ = 989-993 for 1002 degrees of freedom) across all values of $\cos\mu$. For this source also, \(R\) decreased with increasing \(\cos\mu\). The representative values of \(R\) are \(1.22^{+0.51}_{-0.45}\), \(0.52^{+0.21}_{-0.18}\), \(0.30^{+0.12}_{-0.10}\), and \(0.23^{+0.09}_{-0.08}\) for \(\cos\mu\) = 0.1, 0.3, 0.6, and 0.9, respectively.
However, other parameters remained consistent with the values reported in Table~\ref{tab:bestfit}, irrespective of the assumed $\mu$. These results demonstrate the degeneracy between the $R$ and $\mu$, while confirming the robustness of the other parameters, particularly the plasma temperature, which remain unaffected by the assumed value of $\cos\mu$.

\subsection{Time resolved spectroscopy} \label{sec:spin_phase_resolved}

 \begin{figure}
	\includegraphics[width=\columnwidth]{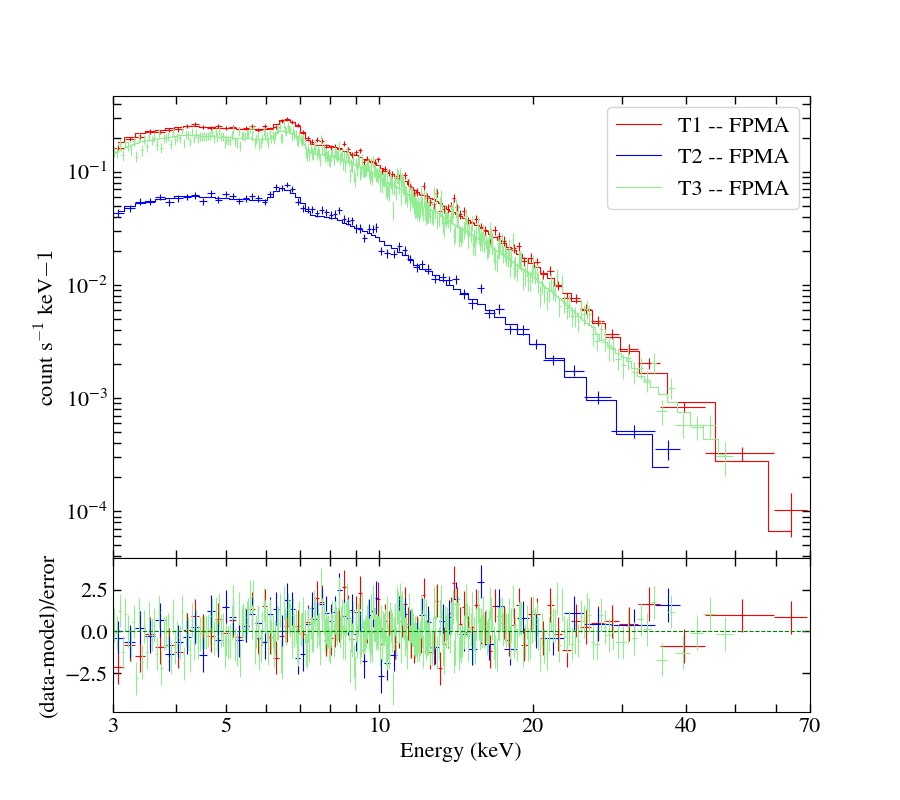}
    \caption{Time-resolved spectra of the source, fitted using the model \texttt{(const$\,*\,$TBabs $*$ (gaussian + reflect $*$ mkcflow))} as described in the text, are shown across three epochs (T1: red, T2: blue, T3: green) in the top panel. The bottom panel shows the corresponding residuals. For clarity, only FPMA data are shown.}
    \label{fig:time_resolved_spectra}
\end{figure}

\begin{table} 
	\centering
	\caption{Best-fit spectral parameters during three time intervals: $\mathrm{T}1 = 0$–$25\,\mathrm{ks}$, $\mathrm{T}2 = 25$–$80\,\mathrm{ks}$, and $\mathrm{T}3 = 80$–$107\,\mathrm{ks}$. The spectra were modeled using the same model used to describe the time averaged spectroscopy \texttt{(const $*$ TBabs(gaussian + reflect(mkcflow)))}.}
    \small
	\label{tab:phaseresolved}
    \setlength{\tabcolsep}{1pt}
	\begin{tabular}{lcccc}
      \hline
    Component & Parameter &   T1 & T2 & T3 \\
      \hline
      Constant         & FPMB               & $1.01\pm0.01$ & $1.02\pm0.02$ & $1.01\pm0.02$\\ 
\texttt{TBabs}                   & $N_{\rm H}$ ($10^{22}\,\rm cm^{-2}$) & $1.59^{+0.67}_{-0.66}$ & $0.11^{+0.97}_{-0.11}$ & $1.54\pm0.66$ \\ 
\texttt{gaussian}  & LineE (keV)     & $6.46^{+0.08}_{-0.10}$ & $6.43^{+0.08}_{-0.09}$ & $6.41\pm0.07$ \\ 
                        & $\sigma$ (keV)     & $0.26^{+0.16}_{-0.14}$ & $0.22^{+0.11}_{-0.12}$ & $0.15^{+0.11}_{-0.12}$ \\ 
                        & Norm ($10^{-5}$) $^{a}$ & $14.07^{+4.98}_{-4.24}$ & $5.06^{+1.07}_{-1.62}$ & $8.79^{+2.86}_{-2.41}$ \\ 
\texttt{mkcflow}  & $kT_{\mathrm{max}}$ (keV)     & $43.68^{+4.75}_{-4.08}$ & $47.44^{+9.54}_{-12.75}$ & $46.12^{+5.86}_{-4.39}$ \\ 
                        & $Z \, (Z_{\sun})$       & $0.56^{f}$ & $0.56^{f}$ & $0.56^{f}$ \\ 
                        & Norm ($10^{-11}$) $^{b}$ & $6.91^{+0.48}_{-0.47}$ & $1.54^{+0.33}_{-0.20}$ & $5.72^{+0.43}_{-0.39}$ \\ 
\texttt{reflect}                 & $R_{\rm refl}$  & $0.51^{+0.22}_{-0.19}$ & $0.21^{+0.59}_{-0.20}$ & $0.28^{+0.18}_{-0.19}$ \\
\hline
\multicolumn{2}{l}{$\chi^2$ (dof)} & $744 (762)$ & $651 (709)$ & $574 (531)$ \\
\hline
   
\end{tabular}
\begin{flushleft}
\noindent
$^{a}$ \texttt{gaussian} normalization (photons $\mathrm{cm}^{-2}\,\mathrm{s}^{-1}$). \\
$^{b}$ \texttt{mkcflow} normalization ($M_{\odot}\,\mathrm{yr}^{-1}$). \\
$^f$ Frozen parameter.
\end{flushleft}
\end{table}

As discussed in Section \ref{sectiming}, the light curve exhibits variability on a timescale of $\sim 85\,\mathrm{ks}$. To investigate the corresponding spectral variability, we divided the $107\,\mathrm{ks}$ observation into three intervals: $\mathrm{T}1 = 0$--$25\,\mathrm{ks}$, $\mathrm{T}2 = 25$--$80\,\mathrm{ks}$, and $\mathrm{T}3 = 80$--$107\,\mathrm{ks}$, which are illustrated by the shaded regions in Fig. \ref{fig_lightcurve} corresponding to two peaks and a broad dip.

For each interval, the FPMA and FPMB spectra were initially fitted jointly using the model
$\texttt{const}*\texttt{TBabs}*(\texttt{mkcflow}+\texttt{gaussian})$.
Since the abundance could not be well constrained due to the limited counts in the time-resolved spectra, we fixed it at 0.56, corresponding to the value obtained from the time-averaged spectral analysis (Table \ref{tab:bestfit}). The fits are statistically acceptable, with $\chi^2/\mathrm{d.o.f.} = 771/763$ for $\mathrm{T}1$, $658/710$ for $\mathrm{T}2$, and $576/532$ for $\mathrm{T}3$.

The normalization of the $\texttt{mkcflow}$ component, which is proportional to the accretion rate $\dot{M}$, is Norm$_1 = 6.26^{+0.41}_{-0.40} \times 10^{-11}\,M_{\odot}\,\mathrm{yr}^{-1}$ (T1), Norm$_2 = 1.46^{+0.15}_{-0.13} \times 10^{-11}\,M_{\odot}\,\mathrm{yr}^{-1}$ (T2), and Norm$_3 = 5.28^{+0.37}_{-0.36} \times 10^{-11}\,M_{\odot}\,\mathrm{yr}^{-1}$ (T3). These normalization values implies that the accretion rates differ significantly among the three intervals. The best-fit maximum plasma temperatures are $57 \pm 4$ keV, $54 \pm 6$ keV, and $55 \pm 4$ keV for $\mathrm{T}1$, $\mathrm{T}2$, and $\mathrm{T}3$, respectively. These temperatures are consistent within uncertainties with each other as well as with the value obtained from the time-averaged spectroscopy.
 
We further included a reflection component in the spectral model for each interval, fixing $\cos\mu = 0.45$. The resulting best-fit spectra and residuals are shown in Fig. \ref{fig:time_resolved_spectra} (for clarity, we have shown only FPMA spectra), and the corresponding best-fit parameters are listed in Table \ref{tab:phaseresolved}. The inclusion of reflection does not alter the results: the mkcflow normalization shows substantial variability between the three intervals (varying by a factor of $\sim4.5$), while no statistically significant change is seen in the maximum plasma temperature.

\section{Discussion} \label{secdiscuss}

The results obtained from the hard X-ray analysis provide constraints on the physical properties of the post-shock accretion region in the short-period asynchronous polar IGR J19552+0044. In this section, we discuss these results and compare them with expectations for mCVs.

\subsection{Detection of reflection in a short period system}

The X-ray spectrum of IGR J19552+0044 is well described by the multi-temperature cooling-flow model (\texttt{mkcflow}) expected for mCVs \citep{mukai2017x}. The detection of significant 6.4 keV Fe K$\alpha$ line with an equivalent width of $161^{+68}_{-65}$ eV and a Compton hump in the 10--50 keV range with a reflection amplitude of $R = 0.37^{+0.15}_{-0.13}$ provide clear evidence for the reprocessing of hard X-rays by a cold, dense material, consistent with reflection from the WD surface, as observed in other long-period mCVs \citep{mukai2015unambiguous}.

The detection of a significant reflection component in IGR J19552+0044 was unexpected for a short-period mCV. Such systems are generally thought to have relatively low specific mass accretion rates, leading to taller post-shock accretion columns that subtend a smaller solid angle at the WD surface and therefore produce weaker Compton reflection and fluorescent Fe K$\alpha$ emission \citep{cropper1999effects,hayashi2014new}. This behavior is well illustrated by the short-period IP EX Hya, for which \citet{luna2018constraining} inferred a tall post-shock region above the WD surface based on the non-detection of a reflection component, with the reflection strength consistent with $R=0$ and an upper limit of $R < 0.15$. Similarly, no statistically significant reflection component has been detected in the other short-period systems observed with NuSTAR, namely CTCV J2056$-$3014 and CD Ind \citep{salcedo2024broadband,dutta2022broad}. However, our results indicate that IGR J19552+0044 exhibits a significant reflection component despite its low mass accretion rate (estimated in Section \ref{disc:sec_constaint_shock}), which implies that the shock forms relatively closer to the WD surface.

The formation of a relatively shorter shock height at a low mass accretion rate in IGR J19552+0044 can be explained if accretion occurs through a smaller footprint on the WD in this AP than is typically expected in mCVs. This results in a higher specific mass accretion rate, and consequently, a shorter shock height \citep{hayashi2014new}. A smaller accretion footprint is expected in APs because stream-fed accretion channels the flow to a localized region near the magnetic pole, whereas accretion in IPs generally produces a more extended arc-shaped footprint on the WD surface \citep{king1984hard,hellier2001cataclysmic}.

\subsection{Constraint on WD mass and shock height} \label{disc:sec_constaint_shock}

Spectral fitting with the cooling-flow model yields a maximum post-shock plasma temperature of $kT_{\rm max}=44.5^{+3.6}_{-3.3}$ keV. In mCVs, the maximum shock temperature is primarily determined by the depth of the WD gravitational potential and therefore provides an important diagnostic of the WD mass \citep{cropper1999effects,mukai2017x}. Measurements of WD mass from hard X-ray spectroscopy are available for many long-period mCVs, but only a handful of short-period systems are known, and even fewer have been reported using X-ray observatories \citep{cropper1998white,suleimanov2019hard,shaw2020measuring,suleimanov2025x}. We estimate the WD mass $M_{\rm WD} = 0.88\pm0.03\,M{_\odot}$ using the standard relation between the shock temperature and the gravitational potential of the WD \citep{mukai2017x}, together with the mass--radius relation for cold white dwarfs \citep{nauenberg1972analytic}. The corresponding WD radius is $R_{\rm WD} = (6.4\pm0.2) \times 10^{8}$ cm. However, if the system's magnetic field is sufficiently strong for cyclotron cooling to become significant, this estimate would correspond to an underestimate of the WD mass \citep{cropper1998white}. A finite shock height can also result in an underestimate of the WD mass \citep{suleimanov2025x}.
Taking these considerations into account, the inferred WD mass should be regarded as a lower limit of $M_{\rm WD}\geq 0.85,M_{\odot}$.

We also placed a constraint on the shock height from the detected reflection component. Assuming an average inclination angle of the reflecting surface $\mu = 63^\circ$, the measured reflection amplitude $R = 0.37^{+0.15}_{-0.13}$ corresponds to a shock height above the WD surface of $H = 0.29^{+0.30}_{-0.15} \, R_{\mathrm{WD}}$, based on the reflection amplitude–shock height relation \citep{tsujimoto2018suzaku}. 
However, as discussed in Section~\ref{sec:spectral_analysis}, $R$ is degenerate with $\mu$. Furthermore, \citet{tovmassian2017igr} reported an upper limit on the inclination angle of this source $72^\circ$, disfavoring $\cos\mu < 0.31$. 
For \(\cos\mu\) = 0.3–0.9, we found that the shock height ranges from \(\sim 0.14\) to \(\sim 0.57\,R_{\mathrm{WD}}\).

The X-ray flux measured in the 3--50~keV energy band is $F_{\rm X} = 1.03 \pm 0.01 \times 10^{-10}\, \mathrm{erg\, cm^{-2}\, s^{-1}}$. The observed flux $F_{\mathrm{X}}$ implies an X-ray luminosity of $L_{\rm X} = 3.42 \pm 0.08 \times 10^{32}\, \mathrm{erg\, s^{-1}}$ for a distance of $166.5\pm1.7$ pc derived from \citet{2023gaia}. The mass accretion rate is estimated from the luminosity and the WD mass and radius using the standard relation between accretion luminosity and gravitational potential energy: 
\begin{equation}
    L_X = \dfrac{G M_{\rm WD} \dot{M}}{R_{\rm WD}}.
    \label{eq:accretionrate_luminosity}
\end{equation}
The inferred mass accretion rate is  \( \dot{M} \sim 3.0 \times 10^{-11} \, M_{\odot} \, \mathrm{yr}^{-1} \), consistent with a low accretion rate system below the period gap \citep{howell2001exploration}.

 Furthermore, we use a simplified theoretical model of the PSR to estimate the fractional accretion area on the WD surface ($f$). The shock height ($H$, in cm) depends on the accretion rate, the fractional accretion area, and the mass and radius of the WD, and can be expressed as
\begin{equation}
    H = \frac{1}{3} \times 9.0 \times 10^8 \, \dot{M}_{16}^{-1} f_{-2} M_{\mathrm{WD}}^{3/2} R_{9}^{1/2}, \label{eq:height}
\end{equation}
\citet{frank2002accretion}, where $\dot{M}_{16} $ represents the mass accretion rate converted to \(10^{16} \, \mathrm{g\,s^{-1}}\), and $f_{-2}$ denotes the fractional accretion area on the WD surface, scaled to units of \(10^{-2}\). The parameters \( M_{\mathrm{WD}} \) and \( R_9 \) refer to the WD mass in $M_{\odot}$ and the WD radius in units of \(10^9\) cm, respectively. 
The shock height of \(H \sim 0.3\,R_{\mathrm{WD}}\) corresponds to a fractional accretion area of \(f \sim 1.8 \times 10^{-3}\). Such a small value of \(f\) indicates that the accretion flow is confined to a small region near the magnetic poles on the WD surface.

\subsection{X-ray evidence of Magnetic valve}

The changing orientation of the WD magnetic field with respect to the ballistic accretion stream from the secondary alters the accretion geometry of the post shock region on the WD surface over the beat cycle. The accretion flow is repeatedly directed to different magnetic poles on the WD, a phenomenon known as pole switching. A complex pole-switching model involving four accretion regions on the WD surface has been proposed to explain the beat-phase-dependent variability observed in the TESS light curve of CD Ind \citep{mason2020tess}. In the TESS observations of BY Cam, \citet{mason2022magnetic} discovered a substantial modulation at half the beat period, with brightness varying by a large factor of $\sim 5$. Pole switching primarily redistributes the accretion flow between different accretion regions and modifies their relative visibility, but it is not expected to produce large gradual variations in the mass transfer rate from the donor. To explain this large modulation over half of the beat period in BY Cam, they proposed a magnetic valve mechanism in which the changing relative orientation of the magnetic fields of the white dwarf periodically enhances or suppresses the mass transfer rate through the inner Lagrangian (L1) point.

For the asynchronous polar IGR J19552+0044, \citet{babina2024photometric} identified a photometric modulation at approximately half the beat period from TESS observations and suggested that it may arise from the operation of a magnetic valve. In our \textit{NuSTAR} observation, we also observed a similar variation in the hard X-ray light curve on a comparable timescale of half the beat period (Section~\ref{sectiming}). While the optical brightness is influenced by viewing geometry, cyclotron beaming, reprocessed emission, and the projected emitting area, the hard X-ray emission arises predominantly from the post-shock accretion column and therefore provides a more direct measure of the mass accretion rate. The hardness ratio of the light curve in two energy bands, 3–10 keV and 10–40 keV (Fig.~\ref{fig_lightcurve}) remains nearly constant throughout the observation, indicating that the observed variability is energy independent and is not accompanied by significant spectral change. Time resolved spectroscopy (Section \ref{sec:spin_phase_resolved}) further shows that the spectra are well described primarily by a change in model normalization, corresponding to an inferred variation in the mass accretion rate ($\dot{M}$) by a factor of approximately 4.5. This $\dot{M}$ variability seen at half of the beat cycle provides the first X-ray evidence for the magnetic valve model in an AP.

\subsection{Stable shock temperature despite changing accretion rate}

Table~\ref{tab:phaseresolved} further shows that there is no significant change in PSR temperature in the three time segments ($kT_{\rm max} \approx 45$ keV). This differs from the standard expectation for accretion shocks with finite shock height, where variations in $\dot{M}$ are expected to produce measurable changes in spectral properties such as temperature and reflection strength \citep{cropper1999effects,hayashi2014new,hayashi2018x}.

To examine this expectation, we initially assume that the fractional accretion area (f) does not change between the three intervals and neglect cyclotron cooling. Under this assumption, Eq.~\ref{eq:height} predicts that the lower mass accretion rate during T2 would increase the shock height by a factor of $\sim4.5$ compared with T1 and T3. According to the relation $\mathrm{T}_{\rm s} \propto (R_{\rm WD}+H)^{-1}$ \citep{mukai2017x,suleimanov2025x}, the shock temperature is expected to decrease with increasing shock height, hence a higher shock height should produce a cooler post-shock region. For example, if the shock temperature of $\approx45$ keV is at a shock height of $H=0.1\,R_{\rm WD}$ during T1, the shock temperature is expected to decrease to $\approx34$ keV when the shock height increases to $H\approx0.45\,R_{\rm WD}$ during T2. However, such a decrease is not evident in the observation, as the best-fit PSR temperature during T2 remains close to 45 keV, and its 90\% confidence lower limit is $\approx35$ keV (Table~\ref{tab:phaseresolved}). Moreover, if the shock height during T1 were larger than $0.1\,R_{\rm WD}$, the corresponding shock height during T2 would be even greater, leading to even lower predicted temperature.

Therefore, one possible explanation for the nearly constant shock temperature despite changes in the accretion rate is that the post-shock region has a relatively small shock height, \(H<0.1\,R_{\rm WD}\), at T1. As mentioned in Section~\ref{disc:sec_constaint_shock}, the shock height can have some dependence on the average inclination angle of the reflecting surface, \(\mu\), such that a low shock height corresponds to a high \(\mu\). However, the inclination angle of the source is constrained to \(i<72^\circ\) based on the absence of eclipses \citep{tovmassian2017igr}, which disfavors the higher values of \(\mu\) required for a lower shock height. Thus, the inference of a relatively low shock height should be treated with caution.

Another possible explanation for the stability of the shock temperature is that the shock height does not change significantly with the accretion rate. This could occur if the fractional accretion area on the WD surface ($f$) also varies over the beat cycle. As the location of the threading region where the ballistic accretion stream coupled with the WD magnetic field changes during the beat cycle \citep{mukai1988MNRAS,mason2022magnetic}, the fractional area on the WD surface ($f$) can vary over the beat cycle \citep{frank2002accretion}. In this case, the decrease in $\dot{M}$ during T2 would not necessarily change the shock height if it were accompanied by a decrease in $f$, as the shock height depends on both $\dot{M}$ and $f$ (Eq.~\ref{eq:height}). Consequently, the nearly constant PSR temperature can be explained if the fractional accretion area decreases approximately in proportion to the mass accretion rate, thereby maintaining a nearly constant specific accretion rate.

Cyclotron cooling may also affect the shock height. For magnetic field strengths \(B > 10\)~MG, the dependence of the shock height on the accretion rate may deviate from that described by Eq.~\ref{eq:height}, as illustrated by \citet{belloni2021breaking}. \citet{tovmassian2017igr} constrained the magnetic field strength to \(B < 20\)~MG for this source. However, since the magnetic field strength is not accurately known, we cannot estimate its contribution to maintaining a stable shock temperature despite changes in the mass accretion rate.

Considering the uncertainties in the measurement of the inclination angle and the magnetic field, the correlated variations between the mass accretion rate and the fractional accretion area provide the most favorable explanation for the stable shock temperature despite changes in \(\dot{M}\). A correlated variation in the fractional accretion area with \(\dot{M}\) would imply a nearly constant specific accretion rate and hence a stable shock height and PSR temperature.

\subsection{Comparison of X-ray properties with other APs}

 Hard X-ray properties of only two other APs have been explored with NuSTAR : CD Ind and V1432 Aql \citep{dutta2022broad,wang2021revisiting}. CD Ind and V1432 Aql have shock temperatures of \(kT_{\rm max} \sim 43\) keV and \(kT_{\rm max} \sim 48\) keV, respectively. The three APs (CD Ind, V1432 Aql and IGR J19552+0044) exhibit similar shock temperatures. Since shock temperature primarily depends on the WD mass \citep{mukai2017x}, the similarity among these systems suggests comparable WD masses ($M_{\rm WD} \approx 0.9 \, M_{\odot}$), while observed mCVs typically span a wider mass distribution of \(\sim 0.4\)--\(1.4\) $M_{\odot}$ with an average mass of $0.77\pm0.02 \, M_{\odot}$ \citep{shaw2020measuring}. A bigger sample of APs needs to be observed with NuSTAR to better understand their mass distribution.

CD Ind and V1432 Aql, exhibit an accretion rate of \(\dot{M} \sim 1.3 \times 10^{-11} \, M_{\odot} \, \mathrm{yr}^{-1}\) and \(\dot{M} \sim 1.3 \times 10^{-10}\, M_{\odot} \, \mathrm{yr}^{-1}\), respectively \citep{dutta2022broad,wang2021revisiting}. The former, with an orbital period $\approx 1.87$ h (below the period gap) has an accretion rate comparable to IGR J19552+0044, whereas the latter, with an orbital period of \(\approx 3.37\,\mathrm{h}\) (above the gap), has significantly more accretion rate. 
The difference in accretion rates among these three APs is broadly consistent with the expected rates from CV evolutionary theory, which suggest that longer-period systems have higher secular mass-transfer rates from their donor stars compared to shorter-period systems \citep{knigge2011evolution,pretorius2014constraints,mukai2023orbital}.

In CD Ind, \citet{dutta2022broad} found Fe K$\alpha$ line equivalent width of $\sim 70$ eV and no significant detection of a reflection component. However, the limited quality of the data does not rule out a moderate reflection component comparable to that detected in IGR J19552+0044. In contrast, the long-period asynchronous polar V1432 Aql exhibits a prominent Compton reflection hump with $R>3$ and strong fluorescent Fe K$\alpha$ emission with equivalent width of $\sim 200$ eV, consistent with the trend that a strong reflection is commonly observed in long-period mCVs \citep{wang2021revisiting,mukai2015unambiguous}.

\section{Summary and Conclusion} \label{secconclu}

Hard X-ray spectroscopy of a short-period asynchronous polar IGR J19552+0044 provides new constraints on the structure of the post-shock accretion region on the WD surface. 
The main findings are summarized below:

\begin{enumerate}

\item Spectral analysis yields a maximum shock temperature of $kT_{\mathrm{s}} = 44.5^{+3.6}_{-3.3}$ keV, implying a lower limit on the white dwarf mass of $M_{\rm WD} \geq 0.85\,M_\odot$. The corresponding mass accretion rate is $\dot{M} \sim 3 \times 10^{-11}\,M_\odot\,\mathrm{yr^{-1}}$. 

\item We report the first detection of hard X-ray reflection in a short-period asynchronous polar. The observed reflection amplitude gives the shock height $H\sim0.3\,R_{\rm WD}$ and the fractional accretion area $f\sim1.8\times10^{-3}$.

\item Long term variability that matches with the beat cycle reveals a factor of $\sim 4.5$ change in the mass accretion rate ($\dot{M}$), strongly indicates the magnetic valve mechanism in this asynchronous polar.

\item Despite these large variations in $\dot{M}$, the shock temperature remains approximately constant. This can be explained by correlated variations in \(\dot{M}\) and the fractional accretion area \(f\), which maintains a nearly stable shock height and consequently a stable PSR temperature.
\end{enumerate}

The detection of reflection in an mCV with low accretion rate suggests that reflection can occur in such systems if the accretion footprint is small enough, a scenario that is plausible in stream-fed accretion. This result motivates further investigation of the other mCVs below the period gap to better understand the diversity of accretion geometries. Additionally, APs provide unique laboratories for studying the interplay between stream-fed accretion, pole switching, and the magnetic field of the white dwarf, motivating coordinated polarimetric, photometric, and long-duration X-ray observations to place stronger observational constraints on these processes.

\begin{acknowledgments}
This work is based on observations obtained with \textit{NuSTAR} through the Guest Observer Program Cycle~9 (PI: Anirban Dutta). \textit{NuSTAR} is a project led by the California Institute of Technology, managed and funded by NASA. The authors thank the \textit{NuSTAR} Operations, Software, and Calibration teams for their support with the execution and processing of these observations. This research has made use of data and software provided by the High Energy Astrophysics Science Archive Research Center (HEASARC), which is a service of NASA's Goddard Space Flight Center.
\end{acknowledgments}

\bibliography{reference}{}
\bibliographystyle{aasjournalv7.1}



\end{document}